# Ultra-thin film and interface analysis of high-k dielectric materials employing Time-Of-Flight Medium Energy Ion Scattering (TOF-MEIS)

D. Primetzhofer[(1)], E. Dentoni Litta[(2)], A. Hallén[(2)], M. K. Linnarsson[(2)] and G. Possnert[(1)]

[(1)] Dept. of Physics and Astronomy, Ion Physics, Uppsala University, P.O. Box 516, SE 751 20, Uppsala, Sweden

[(2)] KTH Royal Institute of Technology, ICT, P.O. Box Electrum 229, SE 164 40 Kista, Sweden

**Abstract**

We explore the potential of Time-Of-Flight Medium Energy Ion Scattering (TOF-MEIS) for thin film analysis and analyze possible difficulties in evaluation of experimental spectra. Issues regarding different combinations of composition and stopping power as well as the influence of channeling are discussed. As a model system high-k material stacks made from ultra-thin films of $HfO_2$ grown on a p-type Si (100) substrate with a 0.5 nm $SiO_2$ interface layer have been investigated. By comparison of experimental spectra and computer simulations TOF-MEIS was employed to establish a depth profile of the films and thus obtaining information on stoichiometry and film quality. Nominal film thicknesses were in the range from 1.8 to 12.2 nm. As a supporting method Rutherford-Backscattering spectrometry (RBS) was employed to obtain the areal density of Hf atoms in the films.

## 1. Introduction

Due to the ongoing miniaturization in electronics and sensor technology, conventional elastic scattering techniques using solid state detectors and employing MeV ions have approached their limits when it comes to establishing depth profiles of elements in materials of interest. To refine analysis, approaches to use different detectors and reduce the ion energy, and thus the relative energy loss per unit path length have been made. Medium-Energy Ion Scattering (MEIS) or High-Resolution Rutherford Backscattering Spectrometry (HR-RBS) making use of magnetic and electrostatic spectrometers (see e.g. [1],[2]) as well as modified solid state detectors [3] have emerged from these efforts. Some of these spectrometers have proven to achieve monolayer resolution [4].

Time-of-Flight Medium-Energy Ion Scattering (TOF-MEIS) is a novel elastic scattering technique with the major focus on near-surface crystallography, if performed with a large position sensitive detector. This permits experiments with extremely low primary currents. The TOF-technique furthermore enables to record simultaneously particles with any final energy, which permits large scattering angles even when particles scattered from atoms with lower atomic numbers are detected. This is different from many magnetic or electrostatic spectrometers compensating to some extend for the somewhat lower energy resolution obtainable in TOF-MEIS. The difference in scattering geometry permits thus the investigation of a wider range of different depth without the spectra suffering extensively from multiple scattering effects. However, the potential of TOF-MEIS for characterization of thin films and especially the quality of sub-surface interfaces is still largely unexplored.

In the present investigation, we study well-defined high quality stacks from $HfO_2$ on $SiO_2$ on Si, in order to investigate which film properties can be readily deduced from TOF-MEIS experiments and for which a more detailed analysis is demanded.

## 2. Sample preparation, experiment and simulations:

Samples were prepared the following way: The $SiO_2$ was grown via $O_3$ oxidization and the $HfO_2$ by atomic layer deposition (ALD). A first thickness estimation for the samples was obtained by ellipsometry measurements. The calibration experiments using RBS were performed employing a 5 MV pelletron tandem-accelerator at the Ångström laboratory at Uppsala University. Spectra were obtained at standard RBS conditions employing 2 MeV $^4He^+$-ions as primary projectiles and detecting backscattered particles under a scattering angle of 170° by a solid state detector.

Fig. 1 shows a typical RBS spectrum recorded for a $HfO_2$ film with a thickness subsequently determined to be 12.2 nm (open symbols). The experimental data is fitted by the SIMNRA simulation package [5] (full lines). Since the backscattered yield from Hf appears as a peak, due to limited system resolution care has to be taken to avoid channelling in the Si-substrate. Channelling would lead to erroneous results, due to wrong calibration of the solid angle and beam current. Only a successful fitting of the RBS spectrum in a wide energy range, without deviations towards higher or lower values, especially at the high energy onset of Si can guarantee the absence of channelling. From these experiments it is possible to deduce the total amount of Hf in the film in atoms/$cm^2$ by a fit to the correspondent peak, and consequently the thickness of the films assuming a density and a stoichiometry of the oxide.

The main features of the employed TOF-MEIS set-up will only be briefly summarized, for details see [6]. The set-up is capable to produce atomic and molecular beams of many different ion species in an energy range from approximately 20 – 180 keV. A moveable large solid-angle (diameter of the active area 120 mm) microchannelplate (MCP) detector permits to record position sensitive time-spectra for different scattering angles in backscattering geometry. Due to the large area of the detector the particle flux can be kept extremely low (on the order of 10 nC per spectrum). This permits to use also heavier ions than hydrogen at low ion energies without any significant sample damage due to e.g. sputtering. To deduce information on the thin-film properties, the backscattered yield for a certain narrow scattering angle interval is simulated by Monte-Carlo trajectory calculations by the TRBS-code (Trim for BackScattering) [7]. In the simulations, a correction factor for electronic stopping and different interatomic potentials with tuneable screening (ZBL, TFM [8],[9]) are available. For the present simulations the "universal" ZBL-potential without screening correction was employed.

Fig. 2 presents a typical energy converted experimental spectrum (open symbols) obtained by TOF-MEIS for 45keV $He^+$ ions scattered in an angle of 60 degrees from the film with a nominal thickness (assuming $HfO_2$ and a bulk density of 9.68g/$cm^3$) of 12.2 nm as deduced from RBS (same film as in Fig. 1). The figure also holds different simulations by TRBS (full lines) – details will be explained in the next section.

3. **Results, spectra analysis and discussion:**

The experimental spectrum, converted from TOF to energy, shown in Fig. 2 displays several characteristic features. At the highest energies it still resembles an RBS spectrum, with a characteristic high energy onset followed by a plateau. At lower energies, however, an extended contribution from multiple scattering blurs the onset of the spectrum and leads to a strong background contribution for energies below the low energy edge of the single scattering distribution. Only at much lower energies, this background contains a contribution from scattering from the Si in the substrate, contributing about 2/3 to the observed signal at 20 keV. Also shown are different simulated spectra obtained by TRBS. These simulations help to extract information on the quality of the thin film. The blue line depicts a simulation obtained for a $HfO_2$ film containing the appropriate number of Hf atoms as deduced from RBS embedded in a perfectly homogenous film of $HfO_2$. Obviously, the simulation excellently matches the plateau between the high- and low-energy edge. The shape of the plateau is determined by the energy scaling of stopping power and scattering cross sections in the sample as well as a possible concentration gradient of the scattering species within the compound film. Thus the agreement between simulation and experiment indicate a high degree of homogeneity, i.e. a constant stoichiometry within the film.

At the low energy edge, some deviation between the simulation and the experiment is observed. In contrast to the perfect layer used in the simulation represented by the blue (in print: dark grey) line the simulations shown by black and light grey lines represent simulation employing a sample composition which also exhibits the right Hf content, but instead a mixed layer of Hf, Si and O with a thickness of 1.2 and 2.4 nm corresponding to 10% and 20% of the film thickness respectively. This can also be interpreted as a film roughness. A comparison of the simulations with the experimental dataset clearly indicates that the simulation for 10% roughness clearly fits the experimental spectrum best, whereas the 0 and 20% result in clearly worse fits of the low energy edge.

Whereas parameters regarding the film quality, like homogeneity and roughness can be directly quantified from the TOF-MEIS spectrum obtained from the thin film and subsequent simulations, the absolute film stoichiometry is more difficult to assess. The reason is found in the interdependence of stoichiometry and the stopping power in the films. For stoichiometric $HfO_2$ and known Hf content as an input in the simulations there is no space for a variation of the electronic stopping power. The employed electronic stopping cross section $\varepsilon$ (28.65 eV/$10^{15}$atoms·cm$^2$) which results in the best fit is only slightly (about 10%) exceeding electronic stopping extrapolated from the data by Behar et al. [10] in a transmission experiment. To illustrate the effect of a variation in stopping, Fig. 2 also holds two simulations with electronic stopping modified by +/- 10% (dotted lines). However, by use of a different combination of stoichiometry and stopping power in the compound within certain limits agreement with the experimental spectrum can be obtained. Fig. 3 illustrates several possibilities: the figure holds the same experimental dataset as in Fig. 2, but additional simulations, for different stoichiometry, partially with optimized electronic stopping. As an extreme case the light grey line depicts the spectrum one would obtain for a thin film of Hf with a thickness (in atoms/cm$^2$) according to the RBS results and employing electronic stopping as expected for 45 keV He in Hf. Obviously there is no space for matching this simulation to the experiment, without a tremendous increase of electronic stopping by about a factor of 3. The other simulations represent different examples of changed stoichiometry HfO$_x$ in the range from x = 3 to 0.66, for all of which electronic stopping was modified to match the experimental spectrum. Note that all simulations result in a reasonably well fitted signal from the Si substrate, thus it is not possible to decide on this basis alone whether a certain composition is unreasonable. However, although the fits are of equal quality, the physical model behind becomes unreasonable already before x becomes smaller than 1.5 for which the value of the compound stopping cross section (in eV/10$^{15}$ atoms·cm$^2$) exceeds the value for bulk Hf. Since the stopping of O is found to be much lower than for Hf, the compound stopping is expected to be lower, even if Braggs additivity rule [11] cannot be fully applied at these low ion energies [12]. Reference stopping data thus sets a limit for too strong variation of the composition, an uncertainty in x on the order of 10% however cannot be directly excluded from a single spectrum as the one presented.

Different approaches are typically employed to address this issue in case of a substrate like Si: By choosing the right scattering conditions regarding projectile ion and geometry, the signal for scattering from oxygen can result in a distinct feature in the spectrum allowing for a direct quantification. This approach favours light ions (H,D), for which scattering from oxygen results in a sufficiently high kinematic factor, which permits to separate the signal sufficiently well from the background. This additionally requires a long path length in the sample, either by rather grazing trajectories, or by sufficiently thick samples. As an alternative for thin films one can make use of channelling in the substrate in order to reduce the contribution from Si, resulting in a more pronounced O signal [13].

Additionally, an investigation by different projectiles and/or at different ion energies helps to avoid artefacts from this interdependence of stopping powers and compositions. Also investigating different film thicknesses produced in the same way can indicate inaccuracies in the modelling.

When investigating even thinner films, channelling in the substrate becomes also apparent for He ions, as it is shown in Fig. 4. Experimental data in Fig. 4 clearly illustrates the high energy resolution that can be achieved by TOF-MEIS. In the present case the relative energy resolution $\Delta E/E$ is found better than the value achieved by conventional RBS (about 0.01). The experimental spectrum exhibits a peak structure at the high energy edge of the contribution by scattering from Si. Again, the part of the spectrum originating from Hf indicates very high quality of the employed film, for which the nominal thickness was determined to be 1.8 nm by RBS. When using an electronic stopping power in accordance with the values used for 80 keV He to fit the thicker film (not shown), the width of the spectrum is excellently reproduced. This agreement, in turn, implicates equivalent film composition for the 12.2 and the 1.8nm films.

### 4. Summary and conclusions

The present investigation shows the potential of TOF-MEIS for non-destructive characterization of nm-films. The high-resolution of the technique, together with the ability of recording all energies simultaneously permits to use TOF-MEIS for characterization in a wide range of film thicknesses (about 2 - 20nm). Properties of the thin film, like roughness and homogeneity in composition can be deduced with high precision. To derive film stoichiometry an investigation at a single energy by He ions can yield significant uncertainties in composition. This is linked to uncertainties in compound stopping powers and the dependence of the stopping power on the actual composition. By a set of experiments at different energies and for different film thicknesses or, alternatively if possible, a direct measurement of the yield scattered from oxygen the quality of the information that can be deduced can be improved.

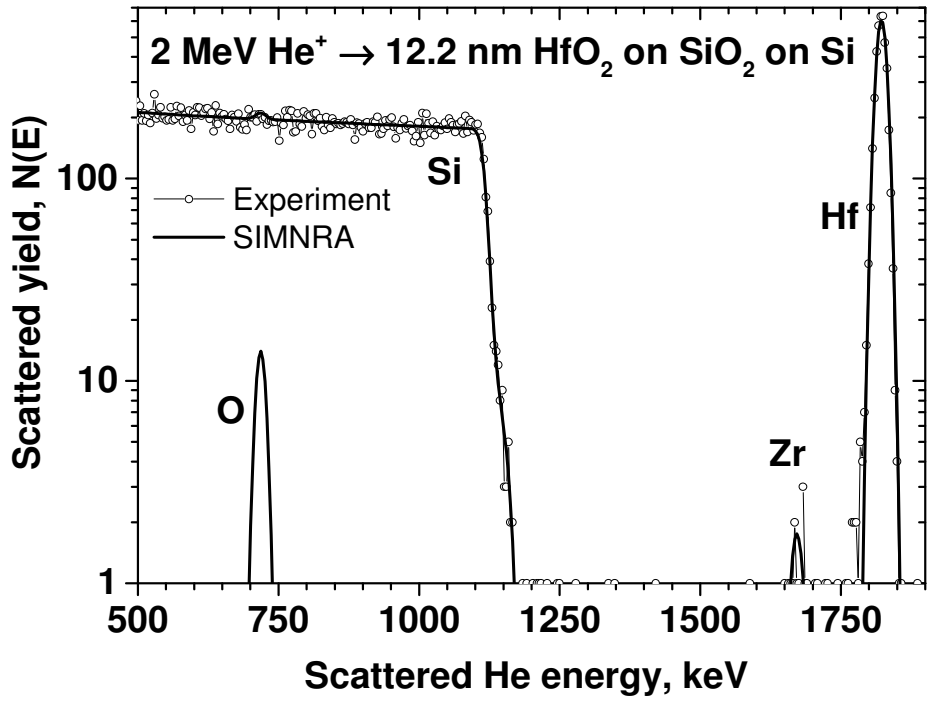

Fig.1: RBS spectrum obtained for 2 MeV $^4$He$^+$ scattered from a HfO$_2$ film on Si with an SiO$_2$ interface layer. The thickness of the film is found to be 12.2 nm when stoichiometric HfO$_2$ composition and bulk density is assumed. A small contribution from Zr (~0.35% atomic per cent assuming the film to be from Hf, O and Zr) can be resolved.

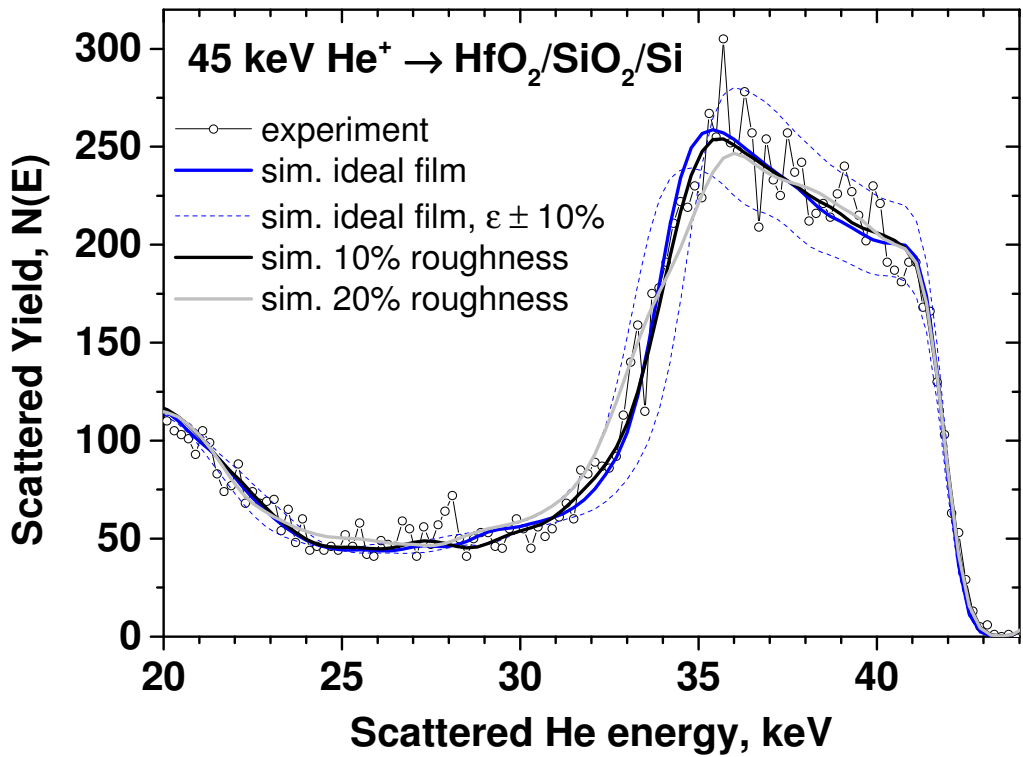

Fig. 2: Energy converted experimental TOF-MEIS spectrum (open symbols) recorded for 45 keV primary $^4$He$^+$ ions scattered from the same film as shown in Fig. 1. The detection angle of scattered particles was 60 degrees. Also shown are TRBS simulations for different roughness in the film (solid lines) as well as a non-appropriate stopping power for HfO$_2$ (dashed lines). For details see text.

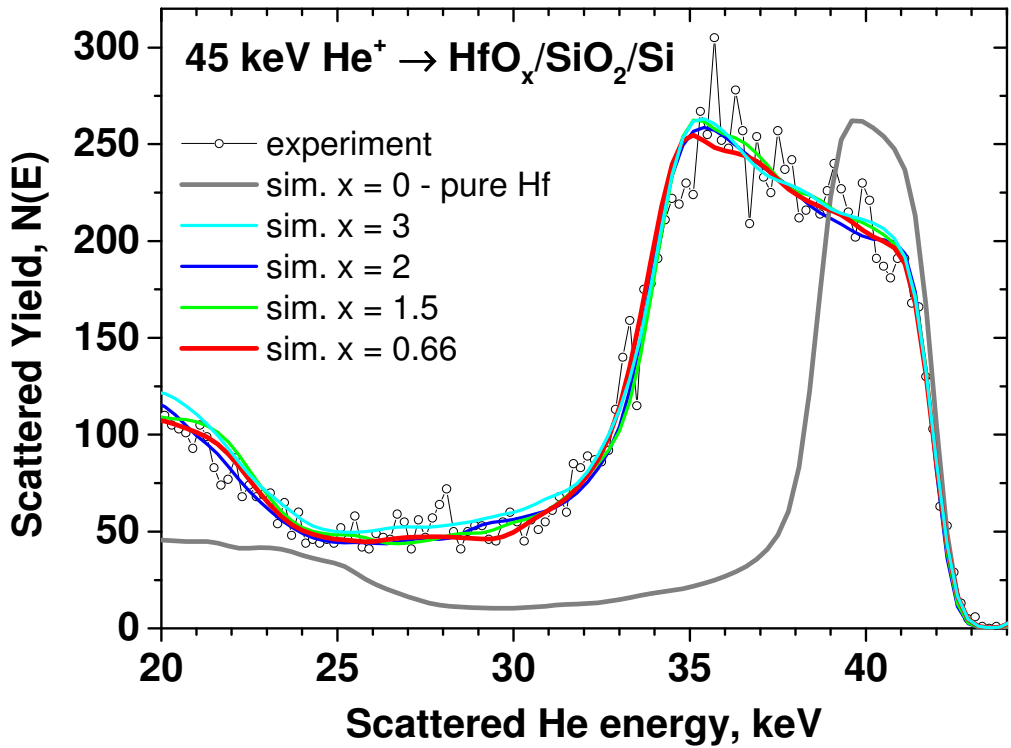

Fig. 3: Energy converted experimental TOF-MEIS spectrum (open symbols) for the same film as in Figs. 1&2. Also shown are TRBS simulations for pure Hf and different composition of the film $HfO_x$ with x between 0.66 and 3 (solid lines) for which electronic stopping has been modified in order to match the experiment.

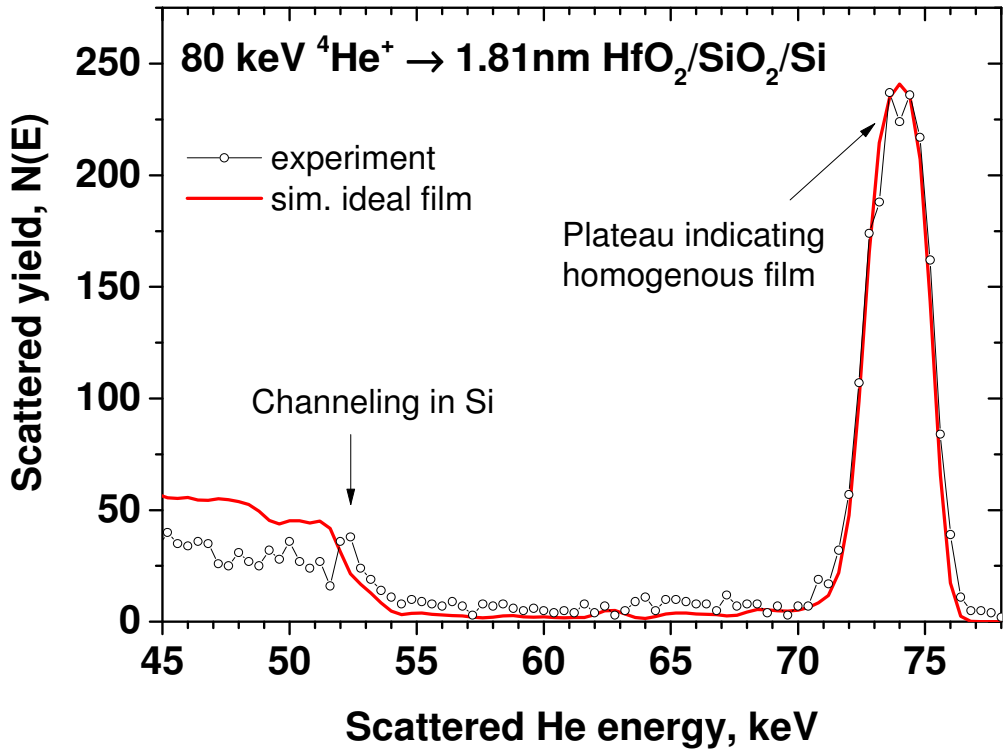

*Fig.4: Energy converted experimental TOF-MEIS spectrum recorded for scattering of 80 keV $^4He^+$ and a detection angle of 105 degrees. The width of the thin film can still be resolved in the spectrum as indicated by the presence of a plateau in the scattered yield. Clear evidence for channelling in the substrate is found in the energy dependence of the scattered yield from Si (surface peak). Also shown is a simulation by TRBS which is found to excellently reproduce the experiment.*